\documentclass[12pt]{article}
\usepackage{epsfig}
\usepackage{a4,isolatin1}
\usepackage{amsmath,amsfonts,latexsym, amssymb}
\newtheorem{satz}{Theorem}[section]
\newtheorem{defi}[satz]{Definition}

\newtheorem{bem}[satz]{Remark}
\newtheorem{lemma}[satz]{Lemma}
\newtheorem{koro}[satz]{Corollary}

\newtheorem{assumption}[satz]{Assumption}

\newtheorem{conclusion}[satz]{Conclusion}
\newtheorem{ob}[satz]{Observation}

\newcommand{\mcal}{\mathcal}

\newcommand{\tit}{\textit}

\newcommand{\R}{\mathbb{R}}

\begin{document}
\thispagestyle{empty}
\begin{center}
\vspace*{1.0cm}

{\LARGE{\bf Scaling Limit and Renormalisation Group in the Critical
    Point Analysis of General (Quantum) Many Body Systems}}

\vskip 1.5cm

{\large {\bf Manfred Requardt }}

\vskip 0.5 cm

Institut f\"ur Theoretische Physik \\
Universit\"at G\"ottingen \\
Bunsenstrasse 9 \\
37073 G\"ottingen \quad Germany\\
(E-mail: requardt@theorie.physik.uni-goettingen.de)

\end{center}

\vspace{1 cm}

\begin{abstract}
  We employ the machinery of smooth scaling and coarse-graining of
  observables, developed recently by us in the context of so-called
  fluctuation operators (inspired by prior work of Verbeure et al) to
  make a rigorous renormalisation group analysis of the critical
  regime. The approach appears to be quite general, encompassing
  classical, quantum, discrete and continuous systems. One of our
  central topics is the analysis of the famous `scaling hypothesis',
  that is, we make a general investigation under what conditions on
  the l-point correlation functions a scale invariant (non-trivial)
  limit theory can be actually attained. Furthermore, we study in a
  rigorous manner questions like the quantum character of the system
  in the scaling limit, the phenomenon of critical slowing down etc.

\end{abstract} \newpage

\section{Introduction}One of the central ideas of the
renormalization group analysis of, say, the critical regime, is
\tit{scale invariance} of the system in the \tit{scaling limit}. This
is the famous \tit{scaling hypothesis} (as to the underlying working
philosophy compare any good text book of the subject matter like e.g.
\cite{Ma} and references therein). Central in this approach is the
socalled \tit{blockspin transformation}, \cite{Ka}. That is,
observables are averaged and appropriately renormalized over blocks of
increasing size. At each intermediate scale a new \tit{effective
  theory} is constructed and the art consists of choosing (or rather:
calculating) the \tit{critical scaling exponents}, so that the
sequence of effective theories converge to a (scale invariant) limit
theory, provided that the start theory lay on the \tit{critical
  submanifold} in the (in general infinite dimensional) parameter
space of theories or Hamiltonians.

Usually the calculations can only be performed in an approximative
way, the main tools being of a perturbative character and being
typically model dependent. Frequently, the more general discussion
concentrates on spin systems to motivate and explain the calculational
steps. While the general working philosophy, based on the concepts of
\tit{asymptotic scale invariance}, \tit{correlation length} and the
like, is the result of a deep physical analysis of the phenomena,
there is, on the other side, no abundance of both rigorous and model
independent results.

This applies in particular to the control of the convergence of the
scaled $l$-point correlation functions to their respective limits if
we start from a microscopic theory, lying on the \tit{critical
  submanifold}. In this case, correlations are typically long-ranged
and the usual heuristic arguments about the interplay between poor
clustering, on the one side, and formation of \tit{block variables} of
increasing size, on the other side, become rather obscure as one is
usually cavalier as to the interchange of various limit procedures.
One knows from examples, that this may be a dangerous attitude in
such a context.

Furthermore, the clustering of the higher correlation functions in the
various channels of phase space may be quite complex and non-uniform
in general.  A concise and selfcontained discussion of the more
general aspects and problems, lurking in the background together with
a useful series of notes and references, can be found in \cite{Pa},
section 7.

Usually, the crucial scaling relation (the \tit{scaling hypothesis})
\begin{equation}W_l^T(Lx_1,\ldots,Lx_l;\mu^*)=L^{-l\cdot n}\cdot
  L^{l\cdot\gamma}\cdot W_l^T(x_1,\ldots,x_l;\mu^*)\end{equation}
which is conjectured to hold at the fixed point (denoted by $\mu^*$ in
the parameter space), is the starting point (or physical input) of the
analysis. Here, $W_l^T$ denote the truncated $l$-point functions (see
below), $L$ is the diameter of the blocks, $Lx_i$ are the respective
centers of the blocks, $n$ is the space dimension, $\gamma$ the
statistical renormalisation exponent. If it is different from $n/2$,
we have an `\tit{anomalous}' scale dimension.

In the following analysis, one of our aims is a rigorous investigation
of such (and similar) scaling relations for the $l$-point functions,
starting from the underlying microscopic characteristics of the
theory. We will do this in a quite general manner, that is, the
underlying model theory can be \tit{classical} or \tit{quantum},
\tit{discrete} or \tit{continuous}. We try to make only very few and
transparent assumptions . Our strategy is it, to deal only with the
really characteristic (almost model independent) aspects of the
subject matter. Another goal is it,to derive properties of both the
\tit{intermediate} and \tit{limit states, observables, dynamics} etc.,
with particular emphasis on the quantum aspects. As a perhaps
particularly interesting result we mention a rigorous discussion of
the phenomenon called \tit{critical slowing down}.

What regards the general working philosophy, one should perhaps
mention the framework, expounded in e.g. \cite{Bu} in the context of
the analysis of the \tit{ultraviolet behavior} in algebraic quantum
field theory, or, in the classical regime, the approach of e.g. Sinai
(\cite{Sinai}). While our framework also comprises the classical
regime, it is mainly designed to deal with the quantum case.  In so
far, it is an extension of the methods, developed by us in \cite{Re1},
which, on their side, have been inspired by prior work of Verbeure et
al; see the corresponding references in \cite{Re1}.  Recently we
became aware of a nice treatment of the block spin approach in the
quantum regime in the bock of Sewell (\cite{Sewell}), who employs
methods which are different from ours, but are complementing them
(\tit{quantum (non-) central limits}).

The technical analysis of the convergence behavior of the $l$-point
correlation functions is mainly contained in sections 4 and 5 of the
present paper, which represented the core of a previous preprint
version. The aim of this technical analysis is to isolate the critical
assumptions, which have to be made, in order that the physical picture
comes out correctly. Put differently, we show that the general scaling
picture is by no means an automatic consequence of a few general
physical assumptions but depends on a number of critical details of
the behavior of correlation functions.

We have now added section 3, which contains a discussion of a variety
of general concepts and features being of relevance in the
renormalsation process. We mention e.g. some subtleties concerning the
scale invariance of the limit theory, the (re)construction of the
theory and its dynamics from the correlation functions (which is not
entirely trivial, as the underlying observable algebras are constantly
changing under renormalsation; cf. also \cite{Bu}), the emerging
(non)-quantum character of (parts of) the limit theory, the phenomenon
of \tit{critical slowing down}, which is derived in our setting from
the KMS-property of the limit state.

We want to remark that there exists a superficially different
approach, which is more related to the well established concepts of
renormalisation theory in quantum field theory (see, for example,
\cite{Amit}, \cite{Be} or \cite{Zinn}), that is, renormalising
propagators, Green's functions and path integrals and in which scale
invariance is present on a more implicit level.

Due to lack of space, we do not intend (and actually feel unable) to
relate our approach to the more perturbation theoretically oriented
approaches mentioned above.  We think, these different aspects are
complementing each other.

We concentrate our analysis entirely on the hierarchy of correlation
functions which can be used to define the theory. We generate
renormalized limit correlation functions from them which happen to be
scale invariant (in a sense clarified below), thus defining a new
limit theory via a \tit{reconstruction process}. The nature of this
limit depends on the degree of clustering of the original microscopic
correlation functions. We do not openly discuss the flow of, say, the
renormalized Hamiltonians through parameter space as a sequence of
more and more \tit{coarse-grained} effective Hamiltonians. The
characteristics of these renormalised intermediate theories are
however implicitly given by their hierarchy of correlation functions
as was already explained in e.g. \cite{Re1} or \cite{Bu}.

One should therefore emphasize, that this well-known integrating out
or decimation of degrees of freedom, which characterizes the ordinary
approaches is automatically contained in our approach! The effective
time evolution is carried over from the microscopic theory as
described in \cite{Re1} or (in a slightly other context) in \cite{Bu},
see also \cite{Verbeure} and is redefined on each intermediate scale,
thus implying automatically a rescaling of both the time evolution and
the corresponding Hamiltonian; see section 3. In case we work in an
scenario, defined by ordinary Gibbs states, our framework would
exactly yield these effective Hamiltonians.
\section{The Conceptual Framework}
\subsection{Concepts and Tools}
As to the general framework we refer the reader to \cite{Re1}. One of
our technical tools is a modified (smoothed) version of averaging
(modifications of the ordinary averaging procedure are also briefly
mentioned in the notes in \cite{Pa}). Instead of averaging over blocks
with a sharp cut off, we employ a smoothed averaging with smooth,
positive functions of the type
\begin{equation}f_R(x):= f(|x|/R)\quad\text{with}\quad f(s)=
\begin{cases}1 & \text{for $|x|\leq 1$} \\ 0 & \text{for $|x|\geq 2$}
\end{cases}
\end{equation}
Remark: We will see in the following, that the final result is more or
less independent of the particular class of averaging
functions!\\[0.3cm]
We note that this class of scaled functions has a much nicer behavior
under Fourier transformation, as, for example, functions with a sharp
cut off, the main reason being that the tails are now also scaled. We
have
\begin{equation}\hat{f}_R(k)=const\cdot R^n\cdot \hat{f}(R\cdot
  k)\end{equation}
Remark: One might perhaps think that this choice of averaging will
lead to a different limit theory. This is however not the
case. Furthermore, the mathematical differences between the two
approaches, that is, using sharp or smooth and scaled cut off
functions, are relatively subtle and not so apparent. We are
investigating these aspects in \cite{Re2}.\vspace{0.3cm}

Another point, worth mentioning, are the implications of translation
invariance. We have for the correlation functions
\begin{equation}W(x_1,\ldots,x_n)=W(x_1-x_2,\ldots,x_{n-1}-x_n)\end{equation}
The truncated correlation functions are defined inductively as follows
(see \cite{Re1})
\begin{equation}W(x_1,\ldots,x_n)=\sum_{part}\prod_{P_i}W^T(x_{i_1},\ldots,x_{i_k})\end{equation}
The (distributional) Fourier transform reads
\begin{equation}\tilde{W}^T(p_1,\ldots,p_l)=\hat{W}^T(p_1,p_1+p_2,\ldots,p_1+\cdots
p_{l-1})\cdot\delta(p_1+\cdots p_l)\end{equation}
The dual sets of variables are
\begin{equation}      y_i:= x_i-x_{i+1}\;,\;q_i=\sum_{j=1}^i p_j\quad i\leq
  (l-1)\end{equation}

\subsection{The case of Normal Fluctuations}
As in \cite{Re1}, we assume that away from the critical point the
truncated $l$-point functions are integrable, i.e. $\in
L^1(R^{n(l-1)})$, in the difference variables,\\ $y_i:=x_i-x_{i+1}$. As
observables we choose the translates
\begin{equation}A_R(a_1),\ldots,A_R(a_l)\;,\;A_R(a):=R^{-n/2}\cdot\int
A(x+a)f(x/R)d^nx\end{equation}
(where, for convenience, the labels $1\ldots l$ denote also possibly
different observables). We then get (for the calculational details see
\cite{Re1}, the hat denotes Fourier transform, translation invariance
is assumed throughout, the $const$ may change during the calculation
but contains only uninteresting numerical factors):
\begin{multline}\langle A_R(a_1)\cdots A_R(a_l)\rangle^T=const\cdot
  R^{ln/2}\cdot\\\int \hat{f}(Rp_1)\cdots \hat{f}(-R[p_1+\cdots
 + p_{l-1}])\cdot \hat{W}^T(p_1,\ldots,p_{l-1})\cdot
 e^{-i\sum_1^{l-1}p_ia_i}\cdot e^{ia_l\sum_1^{l-1}p_i}\prod dp_i\\
=const\cdot R^{ln/2}\cdot R^{-(l-1)n}\cdot\\\int \hat{f}(p_1')\cdots
\hat{f}(-[p_1'+\cdots+p_{l-1}'])\cdot
\hat{W}^T(p'_1/R,\ldots,p'_{l-1}/R)\cdot e^{-i\sum_1^{l-1}(p'_i/R)a_i}\cdot e^{ia_l\sum_1^{l-1}p'_i/R}\prod dp'_i\end{multline}
We now scale the $a_i$'s like
\begin{equation}a_i:=R\cdot X_i\;,\;X_i\;\text{fixed}\end{equation}
This yields
\begin{multline}\langle A_R(R\cdot X_1)\cdots A_R(R\cdot
  X_l)\rangle^T=\\const\cdot R^{(2-l)n/2}\cdot \int
  e^{-i\sum_1^{l-1}p'_iX_i}\cdot e^{iX_l\sum_1^{l-1}p'_i}\cdot\\
 \hat{f}(p_1')\cdots\hat{f}(-[p_1'+\cdots+p_{l-1}'])\cdot\hat{W}^T(p'_1/R,\ldots,p'_{l-1}/R)\prod dp'_i\end{multline}

As the $\hat{f}$ are of strong decrease and $\hat{W}^T$ continuous and
bounded by assumption, we can perform the limit $R\to\infty$ under the
integral
and get:\\[0.3cm]
Case 1 ($l\ge3$):
\begin{equation}\lim_{R\to\infty} \langle A_R(R\cdot X_1)\cdots A_R(R\cdot
  X_l)\rangle^T=0\end{equation}
Case 2 ($l=2$):
\begin{equation}\lim_{R\to\infty}\langle A_R(R\cdot X_1)A_R(R\cdot
  X_2)\rangle^T=const\cdot\int \hat{W}^T(0)\cdot
  e^{-ip'_1(X_1-X_2)}\cdot\hat{f}(p'_1)\cdot\hat{f}(-p'_1)dp'_1\end{equation}
\begin{conclusion}In the normal regime, away from the critical point,
  where we assumed $L^1$-clustering, all the truncated correlation
  functions vanish in the limit $R\to\infty$ apart from the $2$-point
  function. We hence have a quasi free theory in the limit as
  described in \cite{Re1} or in the work of Verbeure et al (cf. the
  references in \cite{Verbeure})
\end{conclusion}
\subsection{The Relation to the Heuristic Scaling\\ Hypothesis}
In the following sections we develop a rigorous approach to \tit{block
  -spin renormalisation} in the realm of quantum statistical
mechanics, which tries to implement the physically well-motivated but,
nevertheless, to some extent heuristic scaling hypothesis. The
analysis will be performed both in coordinate space and Fourier space.
In this subsection we restrict our discussion to the two-point
correlation function, for which the asymptotic behavior is simpler and
more transparent.\vspace{0.3cm}

Remark: In the rest of the paper we replace the exponent $n/2$ in the
definition of $A_R(a)$ by a scaling exponent $\gamma'$, which will usually be
fixed during or at the end of a calculation. It plays the role of a
\tit{critical scaling exponent}.\\[0.3cm]
Let us hence study the behavior of
\begin{multline}\label{two} \langle A_R(R\cdot X_1)A_R(R\cdot
  X_2)\rangle^T=R^{-2\gamma'}\cdot\int W^T((x_1-x_2)+R(X_1-X_2))\\\cdot
  f(x_1/R)f(x_2/R)dx_1dx_2\\
=R^{-2\gamma'+2n}\int W^T(R[(x_1-x_2)+(X_1-X_2)])\cdot
f(x_1)f(x_2)dx_1dx_2
\end{multline}
We make the physically well motivated assumption that, in the critical
regime, $W^T$ decays
asymptotically like some inverse power, i.e.
\begin{equation}W^T(x_1-x_2)\sim(const+ F(x_1-x_2))\cdot
  |x_1-x_2|^{-(n-\alpha)}\quad 0<\alpha<n\;,\; F(x)\in L^1\end{equation}
for $|x_1-x_2|\to\infty$, $F$ bounded and well-behaved.

From the last line of $(\ref{two})$ we see that, as $f$ has compact
support, we can replace $W^T$, for $(X_1-X_2)\neq 0$ and $R\to\infty$
by its asymptotic expression and get for $R$ large:
\begin{equation}\langle A_R(R\cdot X_1)A_R(R\cdot
  X_2)\rangle^T\approx const\cdot R^{-2\gamma'+2n}\cdot
  R^{-(n-\alpha)}\cdot\int |y+Y|^{-(n-\alpha)}\cdot f\ast f(y)dy\end{equation}
We choose now
\begin{equation}\gamma'=(n+\alpha)/2\end{equation}
and get a limiting behavior (for $R\to\infty$) as
\begin{equation}const\cdot\int |y+Y|^{-(n-\alpha)}\cdot f\ast f(y)dy\end{equation}
with $y=x_1-x_2,Y=Y_1-Y_2$.

We see that in contrast to the general folklore, the limit correlation
functions are not automatically \tit{strictly} scale invariant but
depend in a weak sense on the chosen smearing functions, $f$. This
phenomenon will be discussed in more detail below as it exhibits a
quite interesting and a little bit hidden aspect. Central in the
renormalisation group idea is that systems on the \tit{critical
  surface} (i.e., critical systems) are driven towards a \tit{fixed
  point}, representing a scale invariant theory.  This idea is usually
formulated in an abstract parameter space of, say, Hamiltonians. In
our correlation function approach the fixed point shows its existence
via the scaling properties of the correlation functions, that is
\begin{equation}W^T_2(L\cdot(X-Y);\mu^*)=L^{-2(n-\gamma')}W_2^T(X-Y;\mu^*)\end{equation}
with $\mu^*$ describing the fixed point in the (usually) infinite
dimensional parameter space. We see from the above that this is
asymptotically implemented by our limiting correlation functions, as we have
(with the choice $\gamma=(n+\alpha)/2$):
\begin{equation}W_2^T(X-Y;\mu^*)\sim|X-Y|^{-(n-\alpha)}\end{equation}
in the asymptotic regime. That is, the above scaling limit leads to a
limit (i.e. fixed point) theory, reproducing the asymptotic
behavior of the original (microscopic) theory.

One should however note that in the more general situation of
$l$-point correlation functions we have to expect a more complex decay
behavior and the existence of various channels as varying clusters of
observables move to infinity. These more intricate technical aspects
will be discussed in the second part of the paper. We continue with a
discussion of a bundle of general properties of the intermediate and
scaling limit systems.
\section{Rigorous Results on the (Quantum) System in the Intermediate
  Regime and in the Scaling Limit}
In this section we assume that the theory exists in the scaling limit
provided that the scaling exponents have been appropriately
chosen. Under this proviso we investigate its algebraic and dynamical
limit structure.
\subsection{The Description of the System at Varying Scales}
In algebraic statistical mechanics we describe a system with the help
of an observable algebra, $\mcal{A}$, a state, $\omega$, or
expectation functional, $<\circ>$, a time evolution,
$\alpha_t$. Frequently one also employs the $GNS$-Hilbert space
representation of the theory, introduced by Gelfand, Naimark, Segal
(see e.g. \cite{Bratteli1}). We already gave a brief discussion of
these points in \cite{Re1}. But as the approach of the scaling limit
is quite subtle both physically and mathematically, we would like to
give a more complete discussion of some of the topics in the following.

We begin with fixing the notation and introducing some technical and
conceptual tools. Expectations on the underlying observable algebra,
$\mcal{A}$, at scale ``$0$'', are given by
\begin{equation}\omega(A(1)\cdots A(l))=:\langle A(1)\cdots
  A(l)\rangle\end{equation}
where, for convenience, different indices may denote different elements,
different times etc. The dynamics is denoted by
\begin{equation}\alpha_t(A)=A(t)\;\text{or}\;A_t\;,\;t\in\R \end{equation}
space translations by
\begin{equation}\alpha_x(A)=A(x)\;\text{or}\;A_x\;,\;x\in\R^n \end{equation}
\begin{equation}\alpha_{t,x}(A)=A(t,x)   \end{equation}

Given such a structure, we can construct a corresponding Hilbert space
representation (for convenience, we use the same symbols for the
algebraic elements).
\begin{equation}\omega\to\Omega\;,\;\omega(A(1)\cdots
  A(l))=(\Omega|A(1)\cdots A(l)\Omega)_{GNS
}   \end{equation}
\begin{equation}\alpha_t\to U_t\;,\;\text{with}\;\alpha_t(A)\to
  U_t\cdot A\cdot U_{-t}  \end{equation}
etc.

The averaged or renormalized observables, $A\to A_R$, are a subset of
elements in the original algebra, $\mcal{A}$. We denote the
subalgebra, generated by these elements, by $\mcal{A}_R$ with
$\mcal{A}_R\subset\mcal{A}$. We can decide to forget the finer algebra,
$\mcal{A}$, and define the \tit{algebra on scale} $R$ by:
\begin{defi}We define the system on scale $R$ by
\begin{equation}\omega^{(R)}(A^{(R)}):=\omega(A_R) \end{equation}
\begin{equation}\alpha_t^{(R)}(A^{(R)}):=(\alpha_t(A))^{(R)} \end{equation}
\begin{equation}\alpha_X^{(R)}(A^{(R)}):=(A(RX))^{(R)}    \end{equation}
that is, we define the objects on the lhs implicitly (by
\tit{reconstruction}) via the following correspondence
\begin{equation}\langle A^{(R)}(t_1,X_1)\cdots
  A^{(R)}(t_l,X_l)\rangle_{(R)}:=\langle A_R(t_1,RX_1)\cdots A_R(t_l,RX_l)\rangle  \end{equation}
\end{defi}
Remark: Note the different treatment of time and
space-translations. We will come back to this point (which has
remarkable physical consequences) below in connection with
\tit{critical slowing down}.
\begin{satz}
  From the above we see that on each scale we have a new theory,
  $\mcal{S}^{(R)}$, which we get by reconstruction from the above
  hierarchy of correlation functions, in particular, a new,
  non-isomorphic algebra, $\mcal{A}^{(R)}$, and a corresponding
  $GNS$-Hilbert space representation. We emphasize that the
  coarse-grained dynamics is also physically different (despite the
  similarities on both sides of the above definitions).

If the scaling limit does exist, we have, by the same token, a scaling
limit system denoted by
\begin{equation}\mcal{S}^{\infty}=(\omega^{\infty},\mcal{A}^{\infty},\alpha_t^{\infty},\alpha_X^{\infty})\end{equation}
with
\begin{equation}\langle A^{\infty}(t_1,X_1)\cdots
  A^{\infty}(t_l,X_l)\rangle=\lim_{R\to\infty}\langle
  A_R(t_1,RX_1)\cdots A_R(t_l,RX_l)\rangle     \end{equation}
\end{satz}
The proof is more or less obvious from what we have said
above.\\[0.3cm]
\begin{koro}We generally assume that $\alpha_t$ is strongly continous
  on $\mcal{A}$. By the above identification process we can
  immediately infer that both $\alpha_t^{(R)}$ and $\alpha_t^{\infty}$
  are also strongly continuous on the corresponding algebras,
  $\mcal{A}^{(R)},\mcal{A}^{\infty}$. By the same token, we can infer
  that $\omega^{(R)}$ and $\omega^{\infty}$ are  $KMS$-states at the
  same inverse temperature $\beta$.
\end{koro}
Proof: Note that the original time evolution ``commutes'' with the
scale transformation in the sense described above. This yields the
mentioned result for all finite $R$. We have in particular that for
suitable elements (for the technical details see \cite{Bratteli2})
\begin{equation}\langle B^{(R)}(t)\cdot A^{(R)}\rangle_{(R)}=\langle A^{(R)}\cdot B^{(R)}(t+i\beta)\rangle_{(R)} \end{equation}
and there exists an analytic function, $F_{AB}^{(R)}$(z), in the strip
$\{z=t+i\tau,\;0<\tau<\beta\}$ with continuous boundary values at
$\tau=0,\beta$:
\begin{equation}F_{AB}^{(R)}(t)= \langle A^{(R)}\cdot B^{(R)}(t)\rangle_{(R)}\;,\; F^{(R)}_{AB}(t+i\beta)= \langle A^{(R)}\cdot B^{(R)}(t+i\beta)\rangle_{(R)}    \end{equation}
This is equivalent to the following equation (cf. \cite{Bratteli2}):
\begin{equation}\int\omega^{(R)}(A^{(R)}\cdot B^{(R)}(t))\cdot f(t)dt=
  \int\omega^{(R)}(B^{(R)}(t)\cdot A^{(R)})\cdot f(t+i\beta)dt    \end{equation}
for $\hat{f}\in \mcal{D}$. As $f(t+i\beta)$ is of strong decrease in
$t$ the limit $R\to\infty$ can be performed under the integral and we
get the same relation in the scaling limit. The above mentioned
equivalence of this property with the $KMS$-condition shows that the
limit state is again $KMS$. This proves the statement.\\[0.3cm]
Remarks: i) Note what we have already said in \cite{Re1}. One reason for
the non-equivalence of the algebras on different scales stems from the
observation that, in general,
\begin{equation}A_R\cdot B_R\neq (A\cdot B)_R\end{equation}
Furthermore, in the scaling limit, many different observables of
$\mcal{A}$ converge to the same limit point, for example, all finite
translates of a fixed observable.\\
ii) A corresonding result in a slightly different context was also
proved in \cite{Bu}.
\subsection{The Scale Invariance of the Limit Theory}
We have seen in sect. 2.3 that the scaling limit of the correlation
functions for the block spin observables is not fully scale invariant
but only asymptotically so (while the short range details of the
original microscopic correlations, encoded in the function
$F(x_1-x_2)$, have been integrated out, there remains an integrated
effect of the initial block-function, $f(x)$ ).

This observation runs a little bit contrary to the general folklore,
in which the various limit procedures are frequently interchanged and
identified without full justification. We will exhibit the true
connections between the various expressions in the following.

With $f(x)$ now being a \tit{general} test function of e.g. compact
support, we have from sect. 2.3, making now the dependence on $f$
explicit
\begin{equation}\lim_{R\to\infty}\langle A_{R,f}(RX_1)\cdot  A_{R,f}(RX_2)\rangle=const\cdot\int |y+Y|^{-(n-\alpha)}\cdot f\ast f(y)dy    \end{equation}
with
\begin{equation}A_{R,f}(RX)=R^{-(n+\alpha)/2}\cdot\int A(RX+x)\cdot f(x/R)dx   \end{equation}
We now rewrite the limit correlation function as
\begin{equation}\langle A^{\infty}_f(X_1)\cdot
  A^{\infty}_f(X_2)\rangle=\int \langle \hat{A}^{\infty}(x_1+X_1)\cdot
  \hat{A}^{\infty}(x_2+X_2)\rangle\cdot f(x_1)f(x_2)dx_1dx_2  \end{equation}
that is, we identify
\begin{equation}A_f^{\infty}(X)=\int \hat{A}^{\infty}(x+X)\cdot f(x)dx   \end{equation}
with $ \hat{A}^{\infty}(x)$ now having rather the character of a \tit{field}
or operator valued distribution.

We have that
\begin{equation}\langle \hat{A}^{\infty}(x_1)\cdot
  \hat{A}^{\infty}(x_2)\rangle=:W^{\infty}(x_1-x_2)= const\cdot |x_1-x_2|^{-(n-\alpha)}
   \end{equation}
Corresponding results would hold for the higher correlation functions,
that is, we arrive at
\begin{conclusion}In contrast to the block observables,
  $A_f^{\infty}$, the field, $\hat{A}^{\infty}(x)$, displays the full scale
  invariance.
\end{conclusion}

The field, $\hat{A}^{\infty}(x)$, can, on the other hand, be directly
constructed by means of a related limit procedure, which is however
\tit{not} of block variable type. We start instead with the
\tit{unsmeared} observables and take the scaling limit, $R\to\infty$
\begin{equation}\lim_R\langle\hat{A}_R(RX_1)\cdot
  \hat{A}_R(RX_2)\rangle\;\text{with}\;\hat{A}_R(RX):=R^{(n-\gamma)}\cdot A(RX)     \end{equation}
and $n-\gamma=(n-\alpha)/2$.\\[0.3cm]
Remark: The extra scaling factor, $R^n$, replaces the missing
integration over the test function, the support of which increases
like $R^n$.\vspace{0.3cm}

Performing the same calculations, we see that the above limit is equal to\\
$\langle A^{\infty}(X_1)\cdot A^{\infty}(X_2)\rangle$. We arrive at
the conclusion
\begin{conclusion}The fully scale invariant limit theory is achieved
  by taking the limits
\begin{equation}\lim_R\langle\hat{A}_R(RX_1)\cdots \hat{A}_R(RX_l)\rangle=:W^{\infty}(X_1,\ldots,X_l)   \end{equation}
The same construction holds of course for the intermediate scales; we
define $\hat{A}^{(R)}(X)$ by the following identification
\begin{equation}\langle\hat{A}^{(R)}(X_1)\cdots
  \hat{A}^{(R)}(X_l)\rangle_{(R)}:=R^{l(n-\gamma)}\cdot\langle
  A(RX_1)\cdots A(RX_l)\rangle\end{equation}
and have for the observables, $A_f^{(R)}$, defined above
\begin{equation}A_f^{(R)}(X)=\int \hat{A}^{(R)}(X+x)f(x)dx \end{equation}
(which can e.g. be checked by direct calculation).
\end{conclusion}
\subsection{The (Non)-Quantum Character in the Scaling Limit}
In the present section we have dealt with model independent properties
of the system, living on scale $R$ or $\infty$. It is clear, that, in
principle, the algebras $\mcal{A}$, $\mcal{A}^{(R)}$ or
$\mcal{A}^{\infty}$, may contain classes of observables which have to
be scaled with different critical exponents. This depends on the
details of the models under discussion and, in particular, on the form
of the \tit{joint spectrum}, $\hat{W}(\omega,k)$, of the Fourier
transforms of e.g. 2-point functions in the vicinity of
$(\omega,k)=(0,0)$ . We think, we have to postpone a more
detailed discussion of all the possible different model classes and
concentrate, for the time being, in this subsection on some generic
aspects.

In subsection B of section 3 of \cite{Re1}, we already discussed the
limiting behavior of commutators of scaled observables. In the regime
of \tit{normal} scaling, that is, scale dimension $\gamma=n/2$, we
found that commutators are non-vanishing in the generic case in the
limit. This means that in general the resulting limit theory is
\tit{non-abelian} (but \tit{quasi-free}!). Perhaps a little bit
surprisingly, the situation changes at the \tit{critical point}, where
the scale-dimensions are, typically, greater than $n/2$ for at least
some observables.

We make the same observation as Sewell in \cite{Sewell}, namely,
commutators of certain ``critical'' observables vanish in the scaling
limit, i.e., the corresponding limit observables are loosing (at
least) part of
their quantum character  .\\
[0.3cm] Remark: We think that the observation that fluctuations and
critical behavior at the critical point are typically of a
\tit{thermal} and not of a \tit{quantal} nature, does somehow belong
to the general folklore in the field of critical phenomena, but we are
not aware at the moment that this fact has been widely discussed in
the literature in greater rigor. Corresponding remarks can e.g. be
found in connection with so-called (\tit{temperature-zero})
\tit{quantum phase transitions} in \cite{Sachev} or Vojta in \cite{Physik} and further references given there.\\[0.3cm]
On the other hand, related phenomena were observed in the context of
\tit{spontaneous symmetry breaking} in sect. 6 of \cite{Re1} and for
certain models by Verbeure et al in \cite{Verbeure}. A careful
analysis of the behavior of commutators in a slightly different
context can also be found in \cite{Zagrebnov}.

The general argument goes as follows. We assume that the scaling
exponents for the initial observables, $A,B$, $\gamma_A,\gamma_B$
obey:
\begin{equation}\gamma_A+\gamma_B>n \end{equation}
We then have
\begin{multline}\|[A_R,B_R]\|\leq
  R^{-(\gamma_A+\gamma_B)}\cdot\int\|[A(x_1),B(x_2)]\|\cdot
  f(x_1/R)f(x_2/R)dx_1dx_2\\
=R^{-(\gamma_A+\gamma_B)}\cdot\int\|[A,B(y)]\|\cdot
f(x_1/R)f((x_1+y)/R)dx_1dy
\end{multline}
We assume that the observables $A,B$ are taken at equal times and are
\tit{strictly local}, that is, it exist finite supports
$V_A,V_B\subset \R^n$ so that
\begin{equation}[A,B(x)]=0\;\text{for}\;V_B+x\cap V_A=\emptyset \end{equation}
Remark: The restriction to equal times can be avoided but has then to
be replaced by a cluster assumption on the commutator (see
below).\\[0.3cm]
From the support assumption we immediately infer that the above double
integral is actually a single integral as the commutator on the rhs
vanishes outside a strip of finite diameter. We get
\begin{equation}\lim_R \|[A_R,B_R]\|\leq const\cdot\lim_R
  R^{n-(\gamma_A+\gamma_B)}=0 \end{equation}
as $\gamma_A+\gamma_B>n$ by assumption.

We arrive at the same result if we assume that the above norm of the
commutator happens to be in $L^1(\R^n)$, i.e.
\begin{equation}\|[A,B(y)]\|=:F(y)\in L^1(\R^n) \end{equation}
We have
\begin{multline}R^{-(\gamma_A+\gamma_B)}\cdot\int F(y)\cdot f(x_1/R)f((x_1+y)/R)dx_1dy=
R^{-(\gamma_A+\gamma_B)}\cdot R^{2n}\cdot\int
\hat{F}(p)\hat{f}(Rp)\cdot\hat{f}(-Rp)dp\\
=R^{-(\gamma_A+\gamma_B)}\cdot R^n\cdot\int\hat{F}(p/R)\hat{f}(p)\cdot\hat{f}(-p)dp
\end{multline}
We can again perform the $R$-limit under the integral and get the
limit expression
\begin{equation}R^{n-(\gamma_A+\gamma_B)}\cdot\hat{F}(0)\cdot\int
  \hat{f}(p)\cdot\hat{f}(-p)dp\to 0   \end{equation}
for $R\to\infty$

A simple example where different renormalisation exponents naturally
arise is the following. Take a limit observable, $A^{\infty}(X)$, and
consider its spatial derivative, $\partial_X A^{\infty}(X)$. Then we
have in a slightly sloppy notation (the limit being taken in the sense, described above):
\begin{multline}\partial_X A^{\infty}(X)=\lim_R
  \partial_X(R^{-\gamma_A}\int A(x+RX)\cdot F_R(x)d^nx)\\
=\lim_R (R^{(-\gamma_A+1)}\cdot\int (\partial_x A)(x+RX)\cdot f_R(x)d^nx)
\end{multline}
That is, $\partial_xA=i[P,A]$ has to be scaled with a different scale
exponent. Physically, this can be understood as follows. With
$f_R(x)=f(|x|/R)$ simulating the integration over a ball with radius
$R$, a partial integration in the above formula shifts the
$\partial_x$ to the test function, $f_R(x)$. As $\partial_x f_R(x)=0$
in the interior of the ball, the averaging goes roughly over the
sphere of radius $R$ instead of the full ball. This has to be
compensated by a weaker renormalisation.

Another result in this direction can be found in \cite{Re1} sect.6, in
connection with the canonical Goldstone pair in the context of
spontaneous symmetry breaking.

Further possible candidates are the time derivatives of observables as
e.g. $\langle \dot{A}\dot{A}\rangle$. Fourier transformation yields an
additional prefactor, $\omega^2$ in the spectral weight,
$W_{AA}(\omega,k)$. The $KMS$-condition leads to another constraint:
\begin{equation}W_{AB}(\omega,k)=(1-e^{-\beta\omega})^{-1}\cdot
  W_{[A,B]}(\omega,k)\end{equation}
A combination of such properties shows, that in the scaling limit, the
vicinity of $(\omega,k)=(0,0)$ is important.

From covariance properties (as e.g. in models of relativistic quantum
field theory) one may infer certain characteristics about the
energy-momentum spectrum. For arbitrary models on non-relativistic
many-body theory, the situation is less generic and certainly model
dependent. We refrain from going into the technical details at the
moment.\\[0.3cm]
Remark: We had several discussions with D.Buchholz about this point,
which are gratefully acknowledged. This applies also to the following subsection.

\subsection{The Nature of the Limit Time Evolution and the Phenomenon
  of Critical Slowing Down} We argued above that the appropriate
choice of the respective scaling dimensions of the observables under
discussion is a subtle point and perhaps, to some extent, even a
matter of convenience. After all, one may have some freedom in the
choice of the subset of observables which is to survive the
renormalisation process.

We will not give a complete analysis of all possibilities in the
following but rather emphasize one, as we think, particularly
remarkable phenomenon, namely, the phenomenon of \tit{critical slowing
  down}. As in the preceding discussion, we choose two observables,
$A,B$, with $\gamma_A+\gamma_B>n$, implying that the limit commutator
vanishes. We assume this also to hold for non-equal times, at least
on the level of two-point functions, i.e.
\begin{equation}\langle [A^{\infty},B^{\infty}(t)]\rangle_{\infty}=0  \end{equation}

As the limit state is again a $KMS$-state, the vanishing of the above
commutator implies that the analytic function, $F^{\infty}_{AB}(z)$, fulfills
\begin{equation}F^{\infty}_{AB}(t)=F^{\infty}_{AB}(t+i\beta)     \end{equation}
for all $t$. $F^{\infty}_{AB}(z)$ can hence be analytically continued
to the whole plane and is, furthermore, a globally bounded analytic
function, hence a constant by standard reasoning. We can conclude:
\begin{conclusion}Under the assumptions being made, we have
\begin{equation}\langle A^{\infty}\cdot B^{\infty}(t)\rangle_{\infty}=
const \;\text{for all}\; t\in \R\end{equation}
\end{conclusion}
We see that the subclass of limit observables, which has vanishing
limit commutators (see the preceding subsection), has, by the same
token, time independent limit correlation functions. As these
pair-correlation functions are usually connected with characteristic
observable properties of the system (generalized suszebtibilities,
transport coefficients etc.), this has remarkable physical
consequences. The corresponding phenomenon is called \tit{critical
  slowing down}. For a review see e.g. \cite{Halperin}. In physical
terms, the phenomenon can be understood as follows.

In the critical regime, the patches of strongly correlated degrees of
freedom become very large and extend practically over all scales. That
is, a reorientation of such clusters or a response to external
perturbations takes, if viewed on the microscopic time scale, a very
long time. In the scaling limit this time scale goes to infinity. If
one wants to see observable dynamical effects one must scale the time
also and work with a more macroscopic time scale. We have in the limit
for the unscaled time:
\begin{equation}d/dt\langle A^{\infty}\cdot
  B^{\infty}(t)\rangle_{\infty}=0\end{equation}
This is the same as
\begin{multline}\langle A^{\infty}\cdot[H_{\infty},B^{\infty}(t)]
  \rangle_{\infty}=\lim_R \langle A^{(R)}\cdot[H_{(R)},B^{(R)}(t)]
  \rangle_{(R)}=\\\lim_R d/dt\langle A^{(R)}\cdot  B^{(R)}(t)\rangle_{(R)}
 \end{multline}
(At this place we suppress the discussion of the technical details
connected with the limit processes in order to keep the matter
trasparent).

What one now has to do is obvious. We have to compensate the vanishing
of the above expression in the limit by adding an appropriate scale
factor in the time coordinate. Instead of $B(t)$ we insert
$B(R^{\delta}\cdot t)$ with $\delta$ so chosen that the limit
expression is non-vanishing. Note that differentiation with respect to
$t$ now adds an explicit prefactor $R^{\delta}$. This fixes the
\tit{macroscopic} time scale, $t_m$, for these processes. We can
define
\begin{equation}\langle A^{\infty}\cdot B^{\infty}(t_m)\rangle_{\infty}=
\lim_R\langle A^{(R)}\cdot  B^{(R)}(R^{\delta}\cdot t_m)\rangle_{(R)}
\end{equation}
It is clear that other observables may live on different macroscopic
time scales so that the construction of a common macroscopic limit
time evolution may not be immediate. Such more detailed questions have
to be separately studied for the various model classes.

\section{The Scaling Behavior of the Correlation Functions at the
  Critical Point: Illustration of the Method}
 In the following two
subsections we illustrate, in a first step, the technical methods with
the help of the 2-point functions, which have a more transparent
cluster behavior. A slightly different analysis can already be found
in section 7 of \cite{Re1}. The general idea is it, to extract and
isolate the characteristic singular behavior of the correlation
functions; see also section 2.3 of the present paper. The full
analysis of the cluster behavior of the $l$-point functions is then
given in the following section.
\subsection{\label{One}Method One}
We assume the existence of a certain exponent, $\alpha$, so that
($x^2$ denoting the vector-norm squared) we can make the following
decomposition.
\begin{equation}G(x):=W^T(x)\cdot(1+x^2)^{(n-\alpha)/2}=const+F(x)\end{equation}
with a decaying (non-singular) $F$ which is assumed to be in $L^1$.
Fourier transformation then yields:
\begin{multline}R^{-2\gamma}\cdot\int
  W^T_2((x_1-x_2)+R(X_1-X_2))f(x_1/R)f(x_2/R)dx_1dx_2\\
=R^{-2\gamma}\cdot\int
G((x_1-x_2)+R(X_1-X_2))\cdot[1+((x_1-x_2)+R(X_1-X_2))^2]^{-(n-\alpha)/2}\cdot\\f(x_1/R)f(x_2/R)dx_1dx_2\\
=R^{-2\gamma}\cdot R^{2n-(n-\alpha)}\cdot\int dp\,\hat{G}(p)\cdot
e^{-iRp(X_1-X_2)}\cdot\\
\left[\int e^{-iRp(x_1-x_2)}(R^{-2}+((x_1-x_2)+(X_1-X_2))^2)^{-(n-\alpha)/2}f(x_1)f(x_2)dx_1dx_2\right]\end{multline}
where we made the substitution $x\to R\cdot x$.

We now assume the support of $f$ to be contained in a sufficiently
small ball around zero (or, alternatively, $(X_1-X_2)$ sufficiently
large so that $(x_1-x_2)+(X_1-X_2)\neq 0$ for $x_i$ in the support of
$f$).  With
\begin{equation}\hat{G}(p)=const\cdot\delta(p)+\hat{F}(p)\end{equation}
the leading part in the scaling limit $R\to\infty$ is the
$\delta$-term. Asymptotically we hence get for $R\to\infty$ (setting
$y:=x_1-x_2\;Y:=X_1-X_2$):
\begin{equation}R^{n+\alpha-2\gamma}\cdot const\cdot\int
  |y+Y|^{-(n-\alpha)}\cdot f\ast f(y)dy\end{equation}
with
\begin{equation}f\ast f(y):=\int f(y+x_2)\cdot f(x_2)dx_2\end{equation}
and $y+Y\neq 0$ on $supp(f)$.

The reason why the contribution, coming from $\hat{F}(p)$, can be
neglected for $R\to\infty$ is the following: $f$ is assumed to be in
$\mcal{D}$; by assumption the prefactor never vanishes on the support
of $f(x_i)$. Hence the whole integrand in the expression in square
brackets is again in $\mcal{D}$ and therefore its Fourier transform,
$\hat{g}(p')$, is in $\mcal{S}$ (with $p':=Rp$), that is, of rapid
decrease. We can therefore perform the $R$-limit under the integral
and get a rapid vanishing of the corresponding contribution in $R$ for
 $R\to \infty$.
\begin{equation}\lim_{R\to\infty}R{-n}\cdot\int \hat{F}(p'/R)\cdot
  e^{-ip'Y}\cdot \hat{g}(p')d^np'=0 \end{equation}

As $f\ast f$ has again a compact support, we have that, choosing
\begin{equation}\gamma=(n+\alpha)/2\end{equation}
the limit correlation function behaves as $\sim
|X_1-X_2|^{-(n-\alpha)}$ as in the above heuristic analysis.
\subsection{Method Two}
As in the case of normal clustering or (\cite{Re1}, last section), one
can, on the other hand, improve the too weak decay of $W^T(x_1-x_2)$ and transform it into an integrable (i.e. $L^1$-) function. So, with a
similar notation as in the preceding subsection, we choose a suitable
exponent $\alpha$ in
\begin{equation}P_{\alpha}(x_1-x_2):=(1+|x_1-x_2|^2)^{\alpha/2}\end{equation}
so that
\begin{equation}G(y):=W^T(y)\cdot P_{\alpha}^{-1}(y)\,\in\,L^1\quad
  (y:=x_1-x_2) \end{equation}

In contrast to \tit{Method One}, there is of course a whole range of
such possible exponents, $\alpha>\alpha_{inf}$, so that
\begin{equation}\label{inf}G(y)=\begin{cases}\in L^1 &
    \text{for}\;\alpha>\alpha_{inf}\\
\not\in L^1 & \text{for}\;\alpha<\alpha_{inf}\end{cases}\end{equation}
Proceding as in \tit{Method One}, we get
\begin{multline}R^{-2\gamma}\int W^T(y+R\cdot Y)\cdot
  f(x_1/R)f(x_2/R)dx_1dx_2=\\R^{-2\gamma}\cdot R^{n+\alpha}\int
  dp\,\hat{G}(p/R)\cdot e^{-ipY}\cdot\left[\int
  e^{-ipy}(R^{-2}+(y+Y)^2)^{\alpha/2}\cdot f\ast f(y)dy\right]
\end{multline}

Again the obvious strategy seems to be to choose
\begin{equation}\gamma=(n+\alpha)/2\end{equation}
and perform the limit $R\to\infty$. With the same support properties
as above, that is, $y+Y\neq 0$ for $x_1,x_2\in$ support of $f$, the
integrand in square brackets is again infinitely differentiable with
respect to $y$. Hence, its Fourier transform is fast decaying in
$p$.\\[0.3cm]
Remark: Note that for $\alpha/2$ non-integer and without the above
support restriction, there will show up a singularity in sufficiently
high orders of differentiation for vanishing $R^{-2}$. One can however
control these singularities and show that the analysis still
goes through in the case where the support condition does not
hold. One gets however some mild constraint on the admissible $\alpha$'s.
\\[0.3cm]
Therefore we can again apply Lebesgues' theorem of dominated
convergence and perform the $R$-limit under the integral. This yields
the expression
\begin{multline}\hat{G}(0)\cdot\int dpe^{-ipY}\cdot\left[\int e^{-ipy}\cdot
  |y+Y|^{\alpha}\cdot f\ast f(y)dy\right]=\\
const\cdot \hat{G}(0)\cdot\int \delta(y+Y)\cdot|y+Y|^{\alpha}\cdot f\ast
f(y)dy=
const\cdot\hat{G}(0)\cdot 0\end{multline}
(as a result of the above support condition).
\begin{conclusion}With $\alpha$ chosen so that $G(y)\in L^1$ and
  $\gamma=(n+\alpha)/2$, the limit can be carried out under the
  integral and yields the result zero. This shows a fortiori that
  there is no $\alpha_{min}$ with the property that there is a
  non-vanishing limit-two-point function. Put differently, we have an
  $\alpha_{inf}$ but no $\alpha_{min}$ (cf. (\ref{inf})).
\end{conclusion}

So, in contrast to \tit{Method One}, the relevant exponent,
$\alpha_{inf}$, is of such a peculiar nature that we definitely cannot
apply the above method of interchange of taking the limit $R\to\infty$
and integration. But nevertheless, we will show that
\begin{equation}\gamma:=(n+\alpha_{inf})/2\end{equation}
is the correct critical scaling exponent leading to a sensible limit
theory and that this $\alpha_{inf}$ is exactly the $\alpha$, we have
determined in \tit{Method One}.

We have learned above that in order to arrive at a non-zero limit
correlation function, we are definitely forbidden to exploit
Lebesgues' theorem of dominated convergence in the above
expression. The reason for the vanishing of the respective expression
was that with
\begin{equation}\lim_{R\to\infty}\hat{G}(p/R)=\hat{G}(0)\end{equation}
we have to evaluate $\int \hat{g}(p)dp$ with
\begin{equation}\label{g}\hat{g}(p):=\int e^{-ip(y+Y)}|y+Y|^{\alpha}\cdot f\ast
f(y)dy\end{equation}
This integral happens to be zero due to the explicit factor,
$|y+Y|^{\alpha}$ and the assumed support properties.

So, we have to investigate what happens for $\alpha=\alpha_{inf}$. As
we learned above that there is no $\alpha_{min}$, we can conclude
\begin{ob}For $\alpha=\alpha_{inf}$, $G_{\alpha}(y)$ is no longer in
  $L^1$, with
\begin{equation}G_{\alpha}(y):=W^T(y)\cdot(1+y^2)^{-\alpha/2}\end{equation}
\end{ob}
We know that for $\alpha<\alpha_{inf}$ the decay of $G_{\alpha}(y)$ is
so weak that the Fourier transform develops a power law singularity in
$p=0$; that is, we can conclude
\begin{lemma}For $\alpha_{inf}-\alpha:=\varepsilon$, $\hat{G}_{\alpha}$
  has a singularity of the form $|p|^{-\varepsilon}$ near $p=0$.

For $\alpha=\alpha_{inf}$ the singularity is of logarithmic type near
$p=0$.
\end{lemma}
This statement can again be proved by a scaling argument. Let
$G_{\alpha}$ have a non-integrable tail of the form
$|y|^{-(n-\varepsilon)}$. For the (distributional) Fourier transform we
then have
\begin{equation}\hat{G}_{\alpha}(\lambda\cdot p)=const\cdot\int
  e^{i\lambda py}\cdot G_{\alpha}(y)dy=const\cdot\lambda^{-n}\cdot\int
  e^{ipy'}\cdot G_{\alpha}(y'/\lambda)dy'\end{equation}
For $\lambda\to 0$ we can, as above, replace $G_{\alpha}$ by its
asymptotic expression, which goes as $|y|^{-(n-\varepsilon)}$ and
conclude that $\hat{G}_{\alpha}(\lambda p)$ contains a leading singular
contribution $\sim\lambda^{-\varepsilon}$ (modulo logarithmic
terms). We hence see that
\begin{equation}\hat{G}_{\alpha}(p)\sim|p|^{-\varepsilon}\end{equation}
near $p=0$ as a distribution (that is, the above resoning is to be
understood modulo the smearing with appropriate test functions; see
e.g. \cite{Gelfand}).
For $\alpha=\alpha_{inf}$, the singularity must be weaker than any
power, that is, must be of logarithmic type.

By \tit{Method One} we get a limit correlation function which clusters as
$|Y|^{-(n-\alpha)}$. One may wonder where this behavior is hidden if
we use \tit{Method Two}. Taking only the singular term in
$\hat{G}(p/R)$ into account, we have (with
$\gamma:=(\alpha_{inf}+n)/2$)
\begin{equation}\lim_{R\to\infty}R^{-2\gamma}\langle A_R(RX_1)\cdot
  B_R(RX_2)\rangle^T\sim\lim_{R\to\infty}const\cdot\int \ln
  (|p|/R)\cdot \hat{g}(p)dp\end{equation}
and $\hat{g}(p)$ as in equation (\ref{g}). We can again neglect the term
\begin{equation}\ln R\cdot\int \hat{g}(p)dp\end{equation}
as $\int \hat{g}(p)dp=0$.

Assuming at the moment that $\alpha$ were an integer (we will get the
general result by a scaling argument), the prefactor $|y+Y|^{\alpha}$ can
be transformed into corresponding $p$-differentiations of $\widehat{f\ast
  f}(p)$, which, by partial integration, can then be shifted to
corresponding differentiations of $\ln(|p|)$. This transformation
yields an expression of the type $|p|^{-\alpha}$ times a smooth and
decaying function. That means, we essentially end up with an expression
like
\begin{equation}\int dp\,|p|^{-\alpha}\cdot e^{-ipY}\cdot\left[\int
      e^{-ipy}f\ast f(y)dy\right]\end{equation}
By the same reasoning as above we conclude that the singularity,
$|p|^{-\alpha}$, goes over, via Fourier transform, into a weak decay
proportional to $|X_1-X_2|^{-(n-\alpha)}$, that is, we arrive at the
same result as in \tit{Method One}, whereas the reasoning is a little
bit more tricky.

For a general non-integer $\alpha$ the argument could be made precise by
analysing the distributional character of an expression like
$r^{\beta}$ with $r:=|x|$ and its Fourier transform. As the analysis
is a little bit tedious, we refer the reader to \cite{Gelfand}. On the
other hand, one can use a scaling argument as above (with
$Y:=\lambda\cdot Y_0\,,\,Y_0\;\text{fixed as}\;\lambda\to\infty$). This
yields an asymptotic behavior of the form
\begin{equation}\lambda^{-(n-\alpha)}\cdot\int
  dp\,\ln(|p|)\cdot\left[\int e^{-i\lambda
      p(y+Y_0)}\cdot|y+Y_0|^{\alpha}\cdot f\ast f(\lambda
    y)dy\right]\end{equation}

The evaluation of the integral for $\lambda\to\infty$ can be done as
follows: As $f\ast f$ has compact support, the volume of the support
of $f\ast f(\lambda y)$ shrinks proportional to $\lambda^{-n}$.
Therefore the expression in square brackets scales as
$\sim\lambda^{-n}$.  On the other hand (due to an `uncertainty
principle' argument), its essential $p$-support increases
proportional to $\lambda^n$. That is, the two effects compensate each
other and we have again a large-$Y$ behavior $\sim|Y|^{-(n-\alpha)}$
as before.

We conclude that both methods lead to the same aymptotic scaling
behavior of the renormalized two-point function.
\section{The General Cluster-Analysis at the Critical Point}
We now study the general situation of the presence of some long-range
correlations in the l-point functions. In contrast to the much simpler
situation prevailing in the case of two-point functions, the
clustering may be quite complicated, in particular, the dependence on
the number, $l$, i.e. the number of observables, occurring in the
expressions, may be non-trivial. Therefore, we have to investigate
these aspects in more detail.\\[0.3cm]
Remark: One should note that our, at first glance, rather technical
analysis serves also the purpose to clarify and isolate the frequently
only tacitly made preassumptions concerning the necessary cluster or scaling
behavior of the correlation functions. Put differently, the preceding
and the following analysis may show which assumptions have actually to
be made, in order that the general picture comes out correctly.\vspace{0.3cm}x1

From general principles (see e.g. \cite{Ruelle}) we know that in a
\tit{pure phase} there is always a certain degree of clustering. We
make the slightly stronger assumption that it is in some way of the
kind of an inverse power law at infinity (to be specified below). We
want to study the scaling limit of
\begin{equation}\label{cor}\langle A_R(R\cdot X_1)\cdots A_R(R\cdot
  X_l)\rangle^T\end{equation}
with
\begin{equation}A_R(a):=R^{-\gamma}\cdot\int
A(x+a)f(x/R)d^nx\end{equation}
and an, at the moment, unspecified exponent, $\gamma$.

The above expression can be written as
\begin{equation}\int
  W^T((x_1-x_2)+R(X_1-X_2),\ldots,(x_{l-1}-x_l)+R(X_{l-1}-X_l))\cdot
  \prod_{i=1}^l f(x_i/R)\prod_{i=1}^l dx_i\end{equation}
Fourier transformation yields (with $\hat{W}^T(q_1,\ldots,q_{l-1})$
considered as a distribution on $\mcal{S}(\R^{(l-1)n})$)
\begin{multline}\label{fourcor}const\cdot R^{l(n-\gamma)}\cdot\int
  \hat{W}^T(q_1,\ldots,q_{l-1})\cdot
  e^{-i\sum_{j=1}^{l-1}Rq_jY_j}\cdot\\\left[\int
    e^{-i\sum_1^{l-1}Rp_jx_j}\cdot e^{iRq_{l-1}x_l}\cdot
    \prod_{i=1}^{i=l}
    f(x_i)\prod_{i=1}^{i=l}dx_i\right]\prod_1^{l-1}dp_j\end{multline}
with $Y_j:=X_j-X_{j+1}$ and the wellknown relation between the
$q$-variables and the $p$-variables (see e.g. section 2 or \cite{Re1}). For
calculational or notational convenience we will employ both
sets of variables which are linear combinations of each other.

As $f$ is in $\mcal{D}$ by assumption, the Fourier transform of
\hspace{0.2cm}$\prod f(x_i)$\hspace{0.2cm} is in $\mcal{S}$ and the
function in square brackets is a function of
$(Rp_1,\ldots,Rp_{l-1})\;\text{or}\;(Rq_1,\ldots,Rq_{l-1})$, being
of rapid decrease in either set of variables. As a
consequence, for $R\to\infty$ and at least one $p_j$ being different
from zero, the expression approaches zero faster than any inverse
power (together with all its derivatives).

From this we see that, as $R\to\infty$, the region of possible singular
behavior is located around $(p)_1^{l-1}=0$ or $(q)_1^{l-1}=0$,
implying also $p_l=-\sum_1^{l-1}p_j=0$. We can hence infer that only
the singular behavior of $\hat{W}^T$ in $(q)=0$ will matter in this
limit. As a consequence, it will be our strategy to isolate this
singular contribution in $\hat{W}^T$ and transform it in a certain
explicit scaling behavior in $R$, which can be encoded in some power,
$R^{-\alpha}$, in front of the integral.

The singular behavior of $\hat{W}^T(q)$ at $(q)=0$ is related to the
weak decay of $W^T(y)$ at infinity. The limiting behavior of $W^T(y)$
can, however, not expected to be simple or uniform (at least not in
the generic case) as $(y_1,\ldots,y_{l-1})$ or $(x_1,\ldots,x_l)$ can
move to infinity in many different ways. We may, for example, have that
$(x_i)$ together with all $|x_i-x_j|$ go to infinity or, on the
other side, the variables move to infinity in certain fixed clusters
of finite diameter. The rate of decay of $W^T(y)$ should of course
depend in general on these details. Correspondingly, the singular
behavior of $\hat{W}^T(q)$ in the infinitesimal
neighborhood of $(q)=0$ should depend on the direction in which
$(q)=0$ is approached, that is, the limit may be direction-dependent.

In the light of this general situation we must at first decide, in
which kind of limit we are mainly interested. Inspecting the
expression (\ref{cor}), we actually started from, we choose in a
first step our fixed vectors, $(X_i)$, so that
\begin{equation}X_i-X_j\neq 0\;\text{for all}\;i,j\end{equation}
As a consequence, all distances, $|RX_i-RX_j|$, go to infinity for
$R\to\infty$. As in the preceding section, we can choose the support of
$f$ so small that, with $x_i,x_j\in supp(f)$, we have
\begin{equation}|R(X_i-x_i)-R(X_j-x_j)|\to\infty\end{equation}
In this particular case we may expect a relatively uniform limit
behavior on physical grounds.\\[0.3cm]
Remark: Similar problems occur in quantum mechanical scattering
theory.\vspace{0.3cm}

Under this proviso the following assumption seems to be reasonable.
\begin{assumption}Under the assumption, being made above, we assume
  the following decomposition of $W_l^T(y)$ to be valid: It exists a
  function, $(1+H(y))$, $H(y)$ homogeneous and positive for $y\neq 0$ so
  that
\begin{equation}G(y):=(1+H(y))\cdot W^T(y)=const+F(y)\end{equation}
with $F$ sufficiently decaying at infinity in the channel, indicated
above, i.e. $\{|y_i|\to\infty\;\text{for all}\;i=1,\ldots,l-1\}$ and
\begin{equation}H(Ry)=R^{\alpha'_l}\cdot H(y)\end{equation}
\end{assumption}
\begin{bem}A typical example for $H(y)$ is $\left(\sum
    y_i^2\right)^{\alpha'_l/2}$.
\end{bem}

Fourier transforming $G(y)$, we get
\begin{equation}\hat{G}(q)=const\cdot\delta(q)+\hat{F}(q)\end{equation}
and expression (\ref{fourcor}) becomes (compare the related expression
in \tit{Method One} of the preceding section)
\begin{multline}const\cdot
  R^{l(n-\gamma)}\cdot\int\,\prod_1^{l-1}dp_j\,\hat{G}(q)\cdot
e^{-i\sum q_jY_j}\cdot\\
\left[\int e^{-i\sum_1^{l-1} Rp_jx_j}\cdot e^{iRq_{l-1}x_l}\cdot
  (1+H(Ry+RY))^{-1}\cdot\prod_1^lf(x_j)\cdot\prod_1^ldx_j\right]\end{multline}

By assumption, $H$ is homogeneous of degree $\alpha'_l$. So we can
extract a negative power of $R$, $R^{-\alpha'_l}$, from the expression
in square brackets. Furthermore, we observed above that for
$R\to\infty$ only the vicinity of $q=0$ matters. Finally, by
assumption, the contribution coming from $\hat{F}(q)$ can be neglected
in this limit (compare the corresponding discussion in the subsection
\ref{One}; as a consequence of the assumed support properties, the
expression in square brackets is again strongly decreasing). We hence have
\begin{multline}\lim_{R\to\infty}\langle A_R(R\cdot X_1)\cdots A_R(R\cdot
  X_l)\rangle^T=\lim_{R\to\infty}const\cdot
  R^{(ln-\alpha'_l-l\gamma)}\cdot
  \int\prod_1^{l-1}dq_j\\\delta(q)\cdot  e^{-i\sum_{j=1}^{l-1}Rq_jY_j}\cdot\left[\int
    e^{-i\sum_1^{l-1}Rp_jx_j}\cdot
    e^{iRq_{l-1}x_l}\cdot\left(R^{-\alpha'_l}+H(y+Y)\right)^{-1}\cdot \prod_{i=1}^{i=l}
    f(x_i)\prod_{i=1}^{i=l}dx_i\right]\end{multline}
Remark: We see again the reason for the special choice being made
above as to the support properties of the functions $f(x_i)$, leading to the
result $y_j+Y_j\neq 0$ on the support of $f$. Without this assumption,
we see for our above example, $H(y)=\left(\sum
    y_i^2\right)^{\alpha'_l/2}$, that in the limit, where
  $R^{-\alpha'_l}$ vanishes, we would get a singular contribution at
  points where $y+Y=0$ in the integrand in square brackets. These terms
  would make the following discussion much more tedious.\vspace{0.3cm}

If we now make the choice
\begin{equation}\gamma:=\gamma_l=n-\alpha'_l/l\end{equation}
we arrive at a finite limit expression, depending on the coordinates $(X_i)$:
\begin{equation}\lim_{R\to\infty}\langle A_R(R\cdot X_1)\cdots A_R(R\cdot
  X_l)\rangle^T=const\cdot\int
  H_l(y+Y)^{-1}\cdot\prod_1^lf(x_i)\prod_1^ldx_i\end{equation}
which is a function of the coarse grained difference coordinates
\begin{equation}Y_j=X_j-X_{j+1}\end{equation}
For the $Y_j$ sufficiently large, it is approximately a function
\begin{equation}W_{limit}(Y)\approx const\cdot
  H_l(Y)^{-1}\end{equation}
That is, the renormalized limit correlation functions reproduce the
asymptotic power law behavior of the original microscopic correlation
functions modulo the convolution with the original smearing functions
as has been discussed already above for the two point functions .

For later use we introduce the new scaling exponent, $\alpha_l$, via
\begin{equation}\alpha'_l+\alpha_l=(l-1)n\end{equation}
This implies
\begin{equation}\label{gamma}\gamma_l=(n+\alpha_l)/l\end{equation}
The underlying reason for this choice is that an asymptotic decay,
$\sim r^{-(l-1)n}$, is just the threshold for $W_l^T$ being integrable
or non-integrable (with $r:=\left(\sum y_j^2\right)^{1/2}$).

We have arrived at the following result: We are interested in a
scaling-limit theory for $R\to\infty$. In order to get a non-vanishing
and finite limit theory, we have to choose the scaling exponent for $l=2$ as
\begin{equation}\gamma=\gamma_2=(n+\alpha_2)/2\end{equation}
Furthermore, we have extracted the asymptotic form from the higher
truncated $l$-point functions, $W_l^T(y)$, and have absorbed it in an
explicit scaling factor, $R$ to some power. If the limit theory is to
be finite, the corresponding scaling exponents for $l>2$ have to be
less or equal to zero. This yields unique $\gamma_l$'s as threshold
values.

A corner stone of the philosophy of the renormalisation group is that
the scaling exponents of the scaled observables remain the same,
irrespectively of the degree of the correlation functions in which
they occur. That is, these exponents are fixed by the exponent,
$\gamma_2$, and we have
\begin{equation}\gamma=\gamma_2\geq \gamma_l\end{equation}
(the latter exponent being derived from equation (\ref{gamma})),
in order that the limit correlation functions remain finite.
\begin{conclusion}We have the following alternatives for $R\to\infty$:
\begin{align}\gamma_2>\gamma_l & \Rightarrow & W_{l,R}^T\to
  0\\
\gamma_2=\gamma_l & \Rightarrow & \lim_{R\to\infty}
  W_{l,R}^T\quad\text{is finite and non-trivial}\\
\gamma_2<\gamma_l & \Rightarrow &  W_{l,R}^T\to\infty\end{align}
If $\gamma_2>\gamma_l$ for all $l\geq 3$, the fixed point is gaussian or
trivial. The limit theory is quasi-free. The limit theory is
non-trivial if $\gamma_2=\gamma_l$ for at least some $l\geq 3$. For
$\gamma_2<\gamma_l$ for some $l$, the limit theory does not exist.
\end{conclusion}
\begin{bem}The corresponding analysis can also be done by employing
  \tit{Method Two} (discussed in the preceding section). One can even
  omit the support conditions assumed above. The treatment then
  becomes more involved but the end result is the same. We discuss one
  particular case below.
\end{bem}

To complete the scaling and/or cluster analysis of the truncated
correlation functions, we have to analyze the other channels and the
respective consequences for scaling exponents and cluster
assumptions.

We mentioned several times that without the support condition
\begin{equation}(X_i-X_j)+(x_i-x_j)\neq 0 \end{equation}
for $x_{i,j}\in supp(f)$, the analysis would become more tedious. On
the other side, this assumption is violated if the observables move to
spatial infinity in certain clusters. The extreme case occurs when all
$X_i$ are chosen to be zero, i.e:
\begin{equation}\langle A_R(1)\cdots
  A_R(l)\rangle^T\;,\;R\to\infty\end{equation}
(the indices 1,\ldots,l denote the different observables).
This scenario was already briefly discussed in section 7 of \cite{Re1}
in connection with phase transitions and/or spontaneous symmetry
breaking, which are also typically related to poor spatial clustering.

With the same notations as above we have
\begin{multline}\langle A_R(1)\cdots A_R(l)\rangle^T=const\cdot R^{l(n-\gamma)}\cdot\\\int
  \hat{W}_l^T(q_1,\ldots,q_{l-1})\cdot
  \left[\int
    e^{-i\sum_1^{l-1}Rp_jx_j}\cdot e^{iRq_{l-1}x_l}\cdot
    \prod_{i=1}^{i=l}
    f(x_i)\prod_{i=1}^{i=l}dx_i\right]\prod_1^{l-1}dp_j\end{multline}
Assuming again the existence of a suitable homogeneous function,
$H_l(y)$, in this channel, we get asymptotically two contributions
\begin{equation}\label{term1}const\cdot R^{l(n-\gamma)-\alpha'_l}\cdot\int
 H_l(y)^{-1}\cdot\prod_1^lf(x_i)\prod_1^ldx_i\end{equation}
and
\begin{multline}\label{term2}const\cdot
  R^{l(n-\gamma)-\alpha'_l}\cdot\int\prod_1^{l-1}dq_j\,
  \hat{F}(q_1,\ldots,q_{l-1})\cdot\\
\left[\int
    e^{-i\sum_1^{l-1}Rp_jx_j}\cdot
    e^{iRq_{l-1}x_l}\cdot\left(H_l(y)\right)^{-1}\cdot \prod_{i=1}^{i=l}
    f(x_i)\prod_{i=1}^{i=l}dx_i\right]\end{multline}

The first term has almost the same form as above. But now the function
in square brackets in the second contribution is no longer of strong
decrease as the integrand (considered as a function of $(x)$or $(y)$)
is no longer in $\mcal{D}$ as it will have a singularity in $y=0$. We
can however provide the following estimate on the degree of this
singularity of $H_l^{-1}$ in $y=0$. We assumed throughout in this
section that $W_l^T$ is not integrable at infinity, that is, the
clustering is weak. On the other side, this asymptotic behavior is
exactly encoded in $H_l^{-1}$, as we observed above. The threshold where
integrability goes over into non-integrability for $H^{-1}_l$ is a
behavior
\begin{equation}\sim
  r^{-(l-1)n}\;,\;r:=\left(\sum_1^{l-1}y_j^2\right)1/2\end{equation}

We can therefore conclude that
\begin{equation}\alpha'_l\leq (l-1)n\end{equation}
in the above construction if $W_l^T$ is non-integrable at infinity. If
$\alpha'_l$ is even strictly smaller than $(l-1)n$, which is the ordinary
case in the critical region, we have
\begin{ob}
\begin{equation}\alpha'_l< (l-1)n\end{equation}
implies that $H_l^{-1}$ is integrable near $y=0$. Hence
\begin{equation}H_l^{-1}(y)\cdot \prod_1^l f(x_i)\in L^1\end{equation}
due to the compact support of $f$.
\end{ob}

From this we infer again that, with
\begin{equation}\gamma_l=n-\alpha'_l/l=(n+\alpha_L)/l\end{equation}
the contribution (\ref{term1}) is finite in the scaling limit. For the
contribution (\ref{term2}) we have by the same reasoning that the
function in square brackets is a continuous function of $(Rq)$, which
goes to zero for $Rq\to\infty$ (due to the Riemann-Lebesgue lemma).

On the other side, we have no precise apriori information about $F(y)$
and $\hat{F}(q)$. $F(y)$ goes to zero at infinity as the asymptotic
behavior is contained in $H^{-1}$, but its rate of vanishing is not
clear.
\begin{conclusion}If the integrand of contribution (\ref{term2}) is
  lying in some $L^p$, so that the limit, $R\to\infty$, can be
  performed under the integral, the whole expression vanishes in the
  scaling limit.
\end{conclusion}
In this situation we are left with again with the first term, which
is the limit of
\begin{equation}const\cdot\int
  H_l(y+Y)^{-1}\cdot\prod_1^lf(x_i)\prod_1^ldx_i\end{equation}
for $Y\to 0$. That is, in this case it holds
\begin{satz}If the situation is as in the conclusion,
  $W_l^{lim}(X_1,\ldots,X_l)$ is continuous and we have in particular
\begin{equation}W_l^{lim}(0,\ldots,0)=\lim_{X\to
    0}W_l^{lim}(X_1,\ldots,X_l)\end{equation}
\end{satz}

We can hence resume our findings as follows: If the assumptions, made
above, are fulfilled and if the functions, $H_l$, can be chosen
consistently in all channels, so that the $\gamma_l$'s, resulting from the
relation
\begin{equation}\gamma_l=(n+\alpha_l)/l\end{equation}
are smaller than or identical to $\gamma_2$, we arrive at a full limit
theory, being well-defined in all channels. In this case the
renormalization group program works and yields a non-trivial scaling
limit.\\[1cm]
Acknowledgement: Several discussions with D.Buchholz are greatefully
acknowledged (see also the remark at the end of subsection 3.3).

\end{document}